\documentclass[conference]{IEEEtran}
\IEEEoverridecommandlockouts
\usepackage{cite}
\usepackage{amsmath,amssymb,amsfonts,mathtools}
\usepackage{algorithmic}
\usepackage{graphicx}
\usepackage{textcomp}
\usepackage{xcolor}
\usepackage{float}
\usepackage[pagebackref, colorlinks=true]{hyperref}

\usepackage{comment}
\usepackage{booktabs}
\usepackage{array, makecell}
\usepackage{caption}
\usepackage{fixltx2e}
\usepackage{cleveref}
\usepackage[T1]{fontenc}
\title{A survey on deep learning approaches for breast cancer diagnosis}
\author{\IEEEauthorblockN{
    Timothy Kwong$^{1}$, and Samaneh Mazaheri$^{2}$ \\
    $^1$Faculty of Engineering and Applied Science, \\ Ontario Tech University, Oshawa, ON Canada \\
    $^2$Faculty of Business and Information Technology, \\ Ontario Tech University, Oshawa, ON Canada \\
    \\ timothy.kwong@ontariotechu.net, Samaneh.Mazaheri@ontariotechu.ca} }
\date{}

\begin{document}
\maketitle
\begin{abstract}
Deep learning has introduced several learning-based methods to recognize breast tumours and presents high applicability in breast cancer diagnostics.
It has presented itself as a practical installment in Computer-Aided Diagnostic (CAD) systems to further assist radiologists in diagnostics for different modalities.
A deep learning network trained on images provided by hospitals or public databases can perform classification, detection, and segmentation of lesion types.
Significant progress has been made in recognizing tumours on 2D images but recognizing 3D images remains a frontier so far.
The interconnection of deep learning networks between different fields of study help propels discoveries for more efficient, accurate, and robust networks.
In this review paper, the following topics will be explored: (i) theory and application of deep learning, (ii) progress of 2D, 2.5D, and 3D CNN approaches in breast tumour recognition from a performance metric perspective, and (iii) challenges faced in CNN approaches.
\end{abstract}

\begin{IEEEkeywords}
Mammography, Digital Breast Tomosynthesis, Automatic Breast Ultrasound, MRI, 2D Convolutional Neural Network, 3D Convolutional Neural Network, Classification, Detection, Segmentation
\end{IEEEkeywords}

\section{Introduction}
In 2020, female breast cancer had 2.26 million new cases making breast cancer the highest number of new cases out of 36 cancer sites \cite{sung2021global}.
Moreover, the number of new deaths, due to female breast cancer, was 0.684 million, ranking it the fourth highest of 35 other cancer sites \cite{sung2021global}.
Current modalities for breast cancer screenings include mammography, digital breast tomosynthesis, breast ultrasound, magnetic resonance imaging \cite{american2019breast}. 
Mammography has two types, screen-film mammography and Digital Mammography (DM), where both types are forms of x-ray imaging that use radiation to obtain a 2D image of the breast tissue \cite{american2019breast, canadian2017breast}.
In addition, mammography has facilitated the detection of early stage breast cancer to reduce the risk of cancer death \cite{american2019breast, loberg2015benefits}.
Technological advancements in image acquisition had brought Digital Breast Tomosynthesis (DBT). DBT addressed issues in mammography and delivered improved image acquisition \cite{chong2019digital, canadian2017breast, marinovich2018breast, kopans2014digital}.
It captures multiple 2D images slices of the breast, which are then synthesized into a 3D image \cite{american2019breast, marinovich2018breast, chong2019digital}.
However, these 3D images (volumes) are quasi-3D, due to being a reconstruction of multiple captured 2D images \cite{marinovich2018breast, kopans2014digital}.
Furthermore, image slices are captured using a x-ray tube that pivots parallel to the chest wall along a 15$^{\circ}$ to 60$^{\circ}$ arc \cite{chong2019digital}.
Automatic Breast UltraSound (ABUS) uses high frequency to image the entire breast.
These 2D images are obtained on the transverse plane and synthesized into a 3D volume \cite{giger2016automated}.
Magnetic Resonance Imaging (MRI) uses high-powered magnets along with radio waves generated by a computer to image the breast \cite{american2019breast}.
CAD systems assist radiologists in making diagnostic decisions with higher confidence by providing a "second opinion" \cite{doi2007computer, giger2013breast, cheng2010computer}.
In addition, as mentioned by \cite{van2011computer}, the CAD system should improve radiologists' performance, save time, seamlessly integrate with workflow, and not impose liabilities.
The integration of deep learning algorithms into CAD systems aim to address mentioned objectives above, as well as reducing assessment variability from different radiologists \cite{allison2014understanding, grimm2015interobserver}, reducing recall rate, and increasing cancer detection rate.
The continuous advances in deep learning has brought upon models that outperform radiologists in both classification and localization of cancer tumors in medical images \cite{wang2016deep, mckinney2020international}.
As algorithms advance, computing power becomes more accessible, and expansive well-curated datasets become open-sourced.
Machine learning techniques are able to shift towards state-of-the-art, and aid in tasks within the healthcare sector \cite{ibrahim2020artificial}.
Deep learning, a sub-field of machine learning, conveys representations in simpler forms to solve the problem of learning different representations \cite{Goodfellow-et-al-2016}.
Moreover, deep learning uses multiple interconnected layers of artificial neurons to learn the patterns of simpler expressed forms of the actual representation \cite{singh20203d, Goodfellow-et-al-2016}.
In 2012, a convolutional neural network architecture scored an error rate of 15.3\%, which was 10.9\% lower than the second-best entry \cite{krizhevsky2012imagenet}.
This breakthrough led to an increase in research participation in the field of deep learning, and the continuation of research and usage of CNN architecture for image recognition problems \cite{suzuki2017overview}.
Convolutional Neural Network (CNN) are specialized networks to process data with known grid patterns, as well as learning spatial hierarchies of features within data \cite{Goodfellow-et-al-2016, yamashita2018convolutional}.
As a result, hand-crafted features of cancer tumours are not required for CNNs, considering as CNNs can learn features.
The applicability of deep learning in the medical field presents itself through classification, localization, and segmentation of cancer tumors in medical images from modalities such as MRI, CT, ultrasound.
This review paper will provide insight into deep learning theory, progress of 2D, 2.5D, and 3D CNN architectures, and challenges faced when training a network.

\section{Deep Learning Theory}
This section provides a theoretical overview on deep learning concepts, including data augmentation, building blocks in a typical CNN architecture, overfitting, and transfer learning.

\subsection{Data Augmentation}
Data augmentation is a technique aimed at increasing the dataset size, and improving performance and robustness of the model \cite{gu2018recent, amherd2021heatmap, zheng2016improving}.
Data augmentation methods such as translation, rotation, reflection, blur, and crop, are applied directly on the original image to generate new augmented images.
An instance of data augmentation applies each listed method to an original image to generate 5 new augmented image, which increases the dataset size with new unseen training instances.
Image resizing and greyscaling are other strategies used during data pre-processing to reduce the computation complexity required to process these images.

\begin{equation}\label{eq:1}
    s(t) = (x \ast w)(t) = \sum_{a=-\infty}^{\infty} x(a)w(t-a)
\end{equation}

\begin{figure}
    \centering
    \includegraphics[width=\columnwidth, angle = 0]{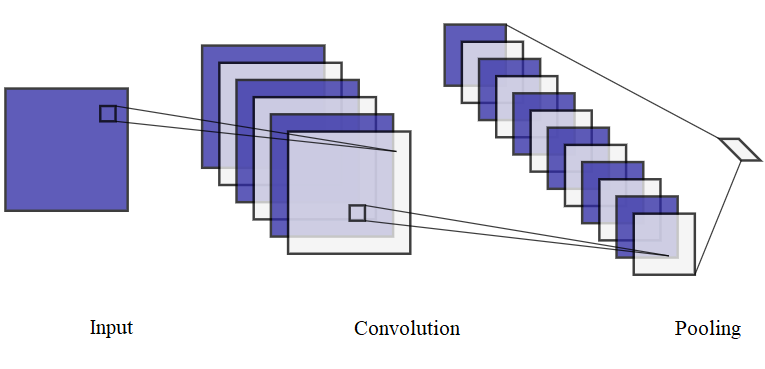}
    \caption{A basic Convolutional Neural Network (CNN) extracts features from an input to output a feature vector. This CNN contains two layers, the convolution and pooling layer. In convolution layer, the entire input is convolved by a kernel, while in the pooling layer, the input is down-sampled. The final output is a flattened column vector containing significant features of the input.
    Adapted from \cite{LeNail2019}. }
    \label{fig1}
\end{figure}

\subsection{2D Convolutional Layer}
Convolutional neural networks use a mathematical operation called convolution; a convolution is a linear operation used, in this case, for feature extraction \cite{Goodfellow-et-al-2016, yamashita2018convolutional}.
Discrete convolution is expressed as equ. (\ref{eq:1}), as seen in \cite[equ. (9.1)]{Goodfellow-et-al-2016}, where s(t) represents the feature map, x(a) represents the input, and w(t-a) represents the kernel.
Moreover, equ. (\ref{eq:1}) illustrates an element-wise product between an input and a kernel to produce a feature map \cite{yamashita2018convolutional, zhang2021dive}.
The 2D convolutional kernel is a matrix of weights that extracts meaningful features from the input for the network to learn and recognize different inputs.
A feature maps can be generated through convolving a kernel with an input, then applying an activation function on the convolved output \cite{gu2018recent}.
In addition, backpropagation is used to update the kernel weights to minimize the loss function \cite{yamashita2018convolutional}.
Stride, as defined by \cite{yamashita2018convolutional}, is "the distance between two successive kernel positions", which dictates the step size of the kernel across the input.
Padding is a technique used to retain the in-plane dimensionality of the feature map even after a convolution further permitting more convolutional layers \cite{yamashita2018convolutional, zhang2021dive}.
Zero-padding aligns the border of the input with zeros to retain the dimensionality \cite{o2015introduction}.
Parameter sharing is a mechanism used in CNN to limit the number of parameters by sharing the kernel weights, which ultimately reduces the model complexity \cite{yamashita2018convolutional, gu2018recent, o2015introduction}.
In addition, parameters can be shared among more abstract features that occur within different images \cite{o2015introduction, wu2017introduction}.
Figure \ref{fig1} illustrates the feature extraction on the input by the convolution and pooling layer to output a feature map.

\subsection{3D Convolutional Layer}
In a 3D CNN, the kernels, stride, and pooling operation are three dimensional, where the third dimension represents a depth dimension \cite{kang20173d}.
This additional dimension allows 3D CNNs to extract features from an additional axis of information.
In 3D convolutional layers, voxel represents the spatial information rather than pixels.

\begin{figure}
    \centering
    \includegraphics[width=0.7\columnwidth, angle = 0]{relu.png}
    \caption{The Rectified Linear Unit (ReLU) function. Adapted from \cite{Hunter:2007}. }
    \label{fig2}
\end{figure}

\subsection{Activation Layer}
The Rectified Linear Unit (ReLU) is an activation function commonly used in neural networks \cite{yamashita2018convolutional, lecun2015deep, ramachandran2017searching, nwankpa2018activation}.
The ReLU function, as shown in the following:
\begin{equation} \label{eq:relu}
    f(x) = max(0, x) = 
    \begin{cases} 
      x, & x \geq 0 \\
      0, & x < 0 \\
   \end{cases}
\end{equation}
and can be seen in \cite[equ. (1.14)]{nwankpa2018activation}, equates values less than zero to zero, and values greater than or equal to zero to passed in value.
A plot of the ReLU function is depicted in Figure \ref{fig2}.

\begin{figure}
    \centering
    \includegraphics[width=0.9\columnwidth, angle = 0]{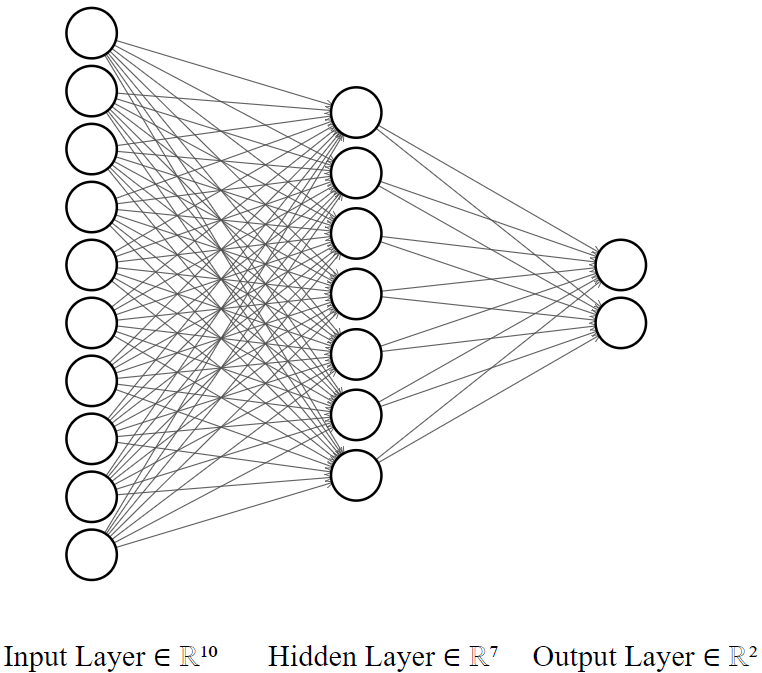}
    \caption{Fully-Connected Neural Network (FCNN).
    Adapted from \cite{LeNail2019}. }
    \label{fig3}
\end{figure}

\subsection{Pooling Layer}
Pooling is a technique to down-sample feature maps \cite{yamashita2018convolutional}, introduces invariance \cite{Goodfellow-et-al-2016}, and merge semantically similar features \cite{Goodfellow-et-al-2016}.
Down-sampling a feature map reduces the in-plane dimensionality \cite{yamashita2018convolutional, sun2017learning}, which reduces the data size without reducing key features in the feature map required for learning.
Max pooling is a pooling operation, which obtains the maximum value within a square region \cite{Goodfellow-et-al-2016, yamashita2018convolutional, sun2017learning}.
Moreover, by obtaining the maximum value, this also makes a representation invariant to small translation or distortions \cite{Goodfellow-et-al-2016, yamashita2018convolutional}.

\begin{figure}
    \centering
    \includegraphics[width=\columnwidth, angle = 0]{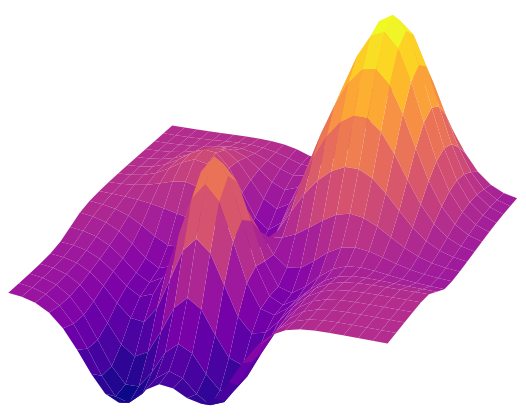}
    \caption{The surface of this 3-dimensional graph represents the objective function. Gradient descent is used to traverse towards the deepest point on this graph. The deepest point represents the global minimum, and parameters that minimize the objective function. Source: Adapted using \cite{Hunter:2007}  }
    \label{fig4}
\end{figure}

\subsection{Fully-Connected Layer} 
The Fully Connected (FC) layer learns a non-linear function to map all the features within a feature space.
Figure \ref{fig3} gives an illustration of the basic structure for a Fully-Connected Neural Network (FCNN).
A CNN can have a FC layer, where the features extracted from the CNN are inputted into the FC layer for a decision output \cite{mazurowski2018deep}.
During training, the goal is to minimize the prediction error made by the CNN, techniques such as back-propagation and gradient descent are used to improve prediction results.
The objective function, also known as the loss function or cost function, determines the difference between the prediction and ground truth; it measures the network error in prediction.
Binary cross-entropy is a loss function used in binary classification.
Back-propagation is a technique used to determine the gradient of the objective function with respect to the weights \cite{lecun2015deep}.
The equation for back-propagation of an objective function with respect to the weight can be calculated using chain rule and is given by the following:

\begin{equation}
    \frac{\partial{L}}{\partial{w}} = \frac{\partial{z}}{\partial{w}} \frac{\partial{a}}{\partial{z}} \frac{\partial{L}}{\partial{w}}
\end{equation}

Gradient descent is used to minimize the objective function through iterative updates of the parameters, such as weights, bias, and kernels, in the negative direction of the gradient of the objective function \cite{yamashita2018convolutional, ruder2016overview}.
The respected equation is given by:

\begin{equation}\label{eq:w}
    w := w - \alpha\frac{\partial{L}}{\partial{w}}
\end{equation}

where w represents the weight, $\alpha$ represents the learning rate, and the partial derivative of the objective function with respect to the weight.
Figure \ref{fig4} illustrates the 3-dimensional surface of an objective function, where the darkest point represents the global minimum of the objective function.

\begin{figure}
    \centering
    \includegraphics[width=7.7cm, angle = 0]{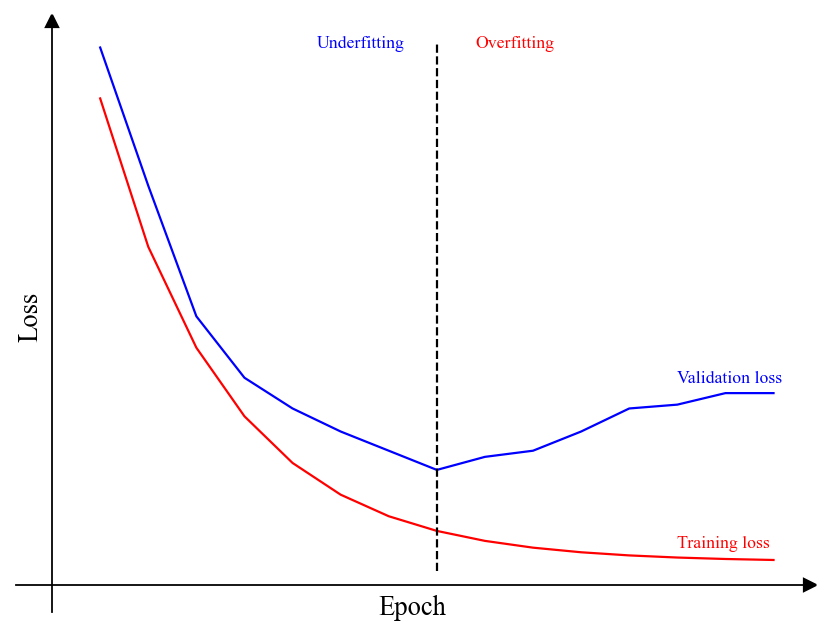}
    \caption{A model that has overfit the training data experiences a divergence in training and validation loss. Adapted from \cite{Hunter:2007}. }
    \label{fig5}
\end{figure}
\subsection{Overfitting}
Models that cannot generalize to new data have overfitting the training data as shown in Figure \ref{fig5}.
Overfitting can be solved through early stopping, network pruning, increasing training data, and regularization \cite{ying2019overview, sarle1996stopped}.
Early stopping is stopping the learning at a point where the curves for training and validation loss are neither overfitting or underfitting \cite{ying2019overview, sarle1996stopped}.
Network pruning involves the pruning of redundant weights, while keeping important weights \cite{liu2018rethinking}.
An increase in training data is required to tune the parameters within a network, so the network can generalize to new data better.
Training data can be increased through collection and/or data augmentation.
Regularization aims to remove useless features and minimize the weights of less important features learned by the model \cite{ying2019overview}.
However, due to the uncertainty of necessary features by the network, a penalty term is added to the objective function to minimize the number of features \cite{ying2019overview}.
Furthermore, there are different types of regularization, such as L1 regularization, L2 regularization (weight decay), and Dropout \cite{ying2019overview}.
Dropout was proposed by Srivastava et al. \cite{srivastava2014dropout} and is another effective strategy in reducing overfitting.

\subsection{Transfer Learning}
Transfer learning is a technique used to improve a learner specialized to a domain through transferring knowledge from a similar domain \cite{weiss2016survey, tan2018survey}.
Additionally, it can be a solution to the issue of insufficient training data \cite{weiss2016survey, tan2018survey, garcia2018survey}.
A quantity of quality training data suitable for effectively training a learner can be expensive, due to difficult collection and curation of data \cite{tan2018survey}.
Henceforth, solution such as homogeneous and heterogeneous transfer learning have been proposed \cite{weiss2016survey}.

\section{Performance Metrics}
A confusion matrix can be used to showcase the different predictions that a classifier makes \cite{davis2006relationship, jeni2013facing}.
A confusion matrix can be visualized as a 2x2 matrix with True Positive (TP), True Negative (TN), False Positive (FP), and False Negative (FN) in one of the grids \cite{jeni2013facing, chicco2020advantages}.
True positives are positive samples correctly predicted as positive, whereas false positives are negative samples incorrectly predicted as positive \cite{davis2006relationship, jeni2013facing}. True negatives are negative samples correctly predicted as negative, whereas false negatives are positive samples incorrectly predicted as negative \cite{davis2006relationship, jeni2013facing}.

The following equations will be used as a means of assessing the deep learning models shown in TABLE \ref{Tab:Tcr1}, \ref{Tab:Tcr3}, and \ref{Tab:Tcr2}. Accuracy is a ratio between the correctly classified and the total samples \cite{jeni2013facing, chicco2020advantages}. The equation for accuracy is given by the following \cite[equ. (2)]{johnson2019survey}:
\begin{equation}\label{eq:3}
    Accuracy = \frac{TP + TN}{TP + TN + FP + FN}
\end{equation}

Sensitivity, also known as recall and true positive rate, is the fraction of relevant (actual true positive) instances that are retrieved \cite{jeni2013facing}. The equation for sensitivity is given by \cite[equ. (5)]{johnson2019survey}:
\begin{equation}\label{eq:4}
    Sensitivity = \frac{TP}{TP + FN}
\end{equation}

Specificity, also known as true negative rate, is the fraction of relevant (actual true negative) instance that are retrieved. The equation for specificity equation is given by \cite[equ. (6)]{johnson2019survey}:
\begin{equation}\label{eq:5}
    Specificity = \frac{\text{TN}}{TN + FP} 
\end{equation}

Precision is the fraction of retrieved instances that are relevant (actual true positive) \cite{jeni2013facing, chicco2020advantages}. The equation for precision is given by \cite[equ. (4)]{johnson2019survey}:
\begin{equation}\label{eq:6}
    Precision=\frac{TP}{TP+FP}
\end{equation}

F score is a weighted ratio measuring the average of precision and recall \cite{jeni2013facing}, the equation for F score is given by \cite[equ. (4)]{chicco2020advantages}:
\begin{equation}\label{eq:7}
    F_1 = 2\: \frac{precision\: \cdot \: recall}{precision+recall}
\end{equation}

Dice Similarity Coefficient (DSC) is a spatial overlap index \cite{zou2004statistical}, and can be used for measuring the segmentation performance of a model. DSC is given by \cite[equ. (2)]{el2019mri}:
\begin{equation}\label{eq:8}
    DSC = \frac{2 \cdot TP}{2\cdot TP + FP + FN}
\end{equation}

Matthews Correlation Coefficient (MCC) calculates for the Pearson product-moment correlation coefficient between the actual and predicted values \cite{chicco2020advantages}. MCC is given by \cite[equ. (2)]{chicco2020advantages}:
\begin{equation}\label{eq:9}
    MCC = \frac{TP \cdot TN - FP \cdot FN}{\sqrt{(TP + FP)(TP + FN)(TN + FP)(TN + FN)} }
\end{equation}

\begin{table*}[t]
    \caption{2D CNN architectures for breast cancer diagnosis}
    \label{Tab:Tcr1}
    \makebox[\textwidth][c]{
        \begin{tabular}{c c c c c c c c}
            \toprule
            & Ref. & Model & Task & Dataset & AUROC (\%) & MCC (\%) & Dice (\%)\\
            \midrule
            DM & \makecell[t]{Dhungel et al. \\(2017) \cite{dhungel2017deep}} & \makecell[t]{CNN +\\ RF +\\Hypothesis-\\ Refinement} &  \makecell[t]{Classification \\Detection \\Segmentation} & \makecell[t]{INbreast} & \makecell[t]{$0.76\pm0.23$ (via RF, \\min. user int.) \\$0.69\pm0.10$ (via CNN,\\ min user int.)} & --- & $0.85\pm0.02$ \\ \\
            & \makecell[t]{Al-antari et al. \\(2018) \cite{al2018fully}} &  \makecell[t]{YOLO +\\ FrCN +\\ CNN} & \makecell[t]{Classification \\Detection \\Segmentation} &  \makecell[t]{INbreast} & \makecell[t]{0.9478 (via CNN)} & \makecell[t]{0.9762 (via YOLO)\\0.8593 (via FrCN)\\0.8991 (via CNN)} & \makecell[t]{0.9269 (via FrCN)}\\ \\
            & \makecell[t]{Chougrad et al. \\(2018) \cite{chougrad2018deep}} &  \makecell[t]{Inception v3} & \makecell[t]{Classification\\Detection} &  \makecell[t]{DDSM \\INbreast \\BCDR} &
            0.99 (via MIAS) & --- & --- \\ \\
            & \makecell[t]{Ribli et al. \\(2018) \cite{ribli2018detecting}} & Faster-RCNN & \makecell[t]{Classification\\Detection} & \makecell[t]{DDSM\\Semmelweis University} & 0.95 (via INbreast) & --- & --- \\ \\
            & \makecell[t]{Singh et al. \\(2020) \cite{singh2020breast}} & cGAN & \makecell[t]{Classification\\Segmentation} & \makecell[t]{DDSM\\INbreast\\Hospital Sant Joan\\ de Reus} & 0.80 & --- & 0.94 \\
            \bottomrule
        \end{tabular}
    }
\end{table*}

\section{Discussion}
This section looks at the different deep learning architectures designed for classification, detection, and segmentation of breast tumours, all are shown in TABLE \ref{Tab:Tcr1}, \ref{Tab:Tcr3}, and \ref{Tab:Tcr2}.

\subsection{Status of Digital Breast Tomosynthesis (DBT)}
DBT has been gaining popularity over digital mammography for higher image quality, richer structural detail, and better reduction in background signal noise \cite{tran2021computational}.
In addition, recent studies \cite{conant2019association, marinovich2018breast, rafferty2016breast, lowry2020screening, conant2016breast} have shown increased cancer detection rate with DBT in women aged 40 to 79 years with dense and non-dense breast, but results on reduced recall rate remains conflicted.
Moreover, \cite{marinovich2018breast, rafferty2016breast, lowry2020screening, conant2016breast} have shown that DBT reduces recall rates for women aged 40 to 79 years with dense and non-dense breast, but \cite{phi2018digital, pattacini2018digital} shown recall rates of DBT plus DM similar to that of DM alone.
DBT reduces the overlapping breast tissue that appear on 2D images as opposed to mammograms \cite{baker2011breast, marinovich2018breast, canadian2017breast}.
As a result, DBT images help with the detection of tumours that may appear overlapped by other healthy breast tissue.
However, a disadvantage of DBT is being less sensitive to imaging malignant calcification and even groups of micro-calcification compared to DM \cite{horvat2019calcifications, tran2021computational}.
In addition, DBT systems that use pixel binning have increased efficiency in detector readings, but in turn reduced 3D spatial resolution \cite{kopans2014digital}.

\subsection{Status of Automatic Breast UltraSound (ABUS)}
The ABUS consists of an ultrasound scanner and a transducer \cite{kim2020automated, shin2015current}.
The ABUS captures axial slices of the breast in different views, then these axial slices are used for 3D reconstruction of sagittal and coronal images \cite{kim2020automated, shin2015current}.

\subsection{Status of Magnetic Resonance Imaging (MRI)} 
Magnetic resonance imaging (MRI) screenings are recommended to patients with a high risk of breast  cancer, due to genetics or family history \cite{lehman2016screening, american2019breast}.
Breast coils are used with an MRI to acquire the image of the breast; the patient lies prone with the breasts in the breast coils before entering the MRI \cite{mann2019breast}.
Breast MRI has different types, such as T1-weighted contrast-enhanced imaging, T2-weighted, ultrafast, and diffusion-weighted imaging \cite{mann2019breast}.
The dimensionality of an acquired image is dependent on the MRI type \cite{greenspan2002mri}.
In 2D image acquisition, multiple 2D image slices of the object are captured, whereas in 3D image acquisition, a true 3D image can be captured \cite{greenspan2002mri, shilling2008super}.

\begin{table*}[t]
    \caption{2.5D CNN architectures for breast cancer diagnosis}
    \label{Tab:Tcr3}
    \makebox[\textwidth][c]{
        \begin{tabular}{c c c c c c c}
            \toprule
            & Ref. & Model & Task & Dataset & AUROC (\%) & Dice (\%)\\
            \midrule
            DM-DBT & \makecell[t]{Yousefi et al. \\(2018) \cite{yousefi2018mass}} & \makecell[t]{DCNN MI-RF-\\based CAD} & \makecell[t]{Classification} & \makecell[t]{Breast Imaging \\Research Laboratory at \\Massachusetts General\\ Hospital (in-house):\\87 DBT images\\(27 malignant,\\60 benign)} & 0.87 & --- \\
            \cmidrule{2-7}
            & \makecell[t]{Kim et al. \\(2017) \cite{kim2017latent}} & \makecell[t]{VGG16 + LSTM} & \makecell[t]{Classification} & \makecell[t]{(in-house)} & 0.919 & --- \\ \\
            & \makecell[t]{Liu et al. \\(2017) \cite{liu20173d}} & \makecell[t]{3D Anisotropic \\Hybrid Network\\(3D AH-Net)} & \makecell[t]{Segmentation} & \makecell[t]{(in-house):\\2809 DBT volumes} & --- & 0.834 \\ \\
            & \makecell[t]{Zhang et al. \\(2020) \cite{zhang20192d}} & \makecell[t]{AlexNet \\(Late fusion +\\Max Pooling)} & \makecell[t]{Classification} & \makecell[t]{(in-house):\\3018 negatives\\272 malignant\\415 benign} & 0.854 & --- \\ \\
            & \makecell[t]{Liang et al. \\(2020) \cite{liang2019joint}} & \makecell[t]{CNN ensemble} & \makecell[t]{Classification} & \makecell[t]{University of\\Kentucky Medical\\Center (in-house):\\DBT and DM\\(709 malignant,\\415 benign)} & 0.97 & --- \\
            \bottomrule
        \end{tabular}
    }
\end{table*}

\subsection{Current Approaches using 2D CNN Architecture }
For 2D classification, Chougrad et al. \cite{chougrad2018deep} adopted state-of-the-art architectures, including ResNet50 \cite{he2016deep}, VGG16 \cite{simonyan2014very}, and Inception v3 \cite{szegedy2016rethinking}, that were pre-trained on ImageNet, and re-purposed for breast cancer screening.
Chougrad et al. \cite{chougrad2018deep} achieved a 0.99 AUC for classification on the MIAS database using a pre-trained and fine-tuned Inception v3 model.
The study concluded that fine-tuning strategy improves classification accuracy on state-of-the-art architecture, and Inception v3 achieved a higher accuracy than VGG16 and ResNet50.

For 2D segmentation, Singh et al. \cite{singh2020breast} adapted upon a study by Isola et al. \cite{isola2017image} to propose a conditional Generative Adversarial Network (cGAN) CAD framework for classification and segmentation breast tumor.
Singh et al. \cite{singh2020breast} had achieved 92.11\% dice coefficient score and 84.55\% IoU for a tight frame of the tumor Region Of Interest (ROI) on cGAN.
Furthermore, the study tested Single Shot Detector (SSD), You Only Look Once (YOLO), and Faster-RCNN and found that SSD achieved the best results on detecting small tumor regions and achieved an overall accuracy of 97\%.
Major contributions proposed in this study include the first adapted cGan for breast tumor segmentation, a multi-class CNN for predicting four breast tumor shapes, and the proposed model outperforming state-of-the-art architecture.

\begin{table*}[t]
    \caption{3D CNN architectures for breast cancer diagnosis}
    \label{Tab:Tcr2}
    \makebox[\textwidth][c]{
        \begin{tabular}{c c c c c c c c c c}
            \toprule
            & Ref. & Model & Dataset & Acc(\%) & AUROC(\%) & AP(\%) & FNR(\%) & Dice(\%) & Fscore(\%)\\
            \midrule
            DBT & \makecell[t]{Zhang et al. \\(2018)\cite{zhang2018classification}} & 3D-T2-Alex & \makecell[t]{University of \\Kentucky (in-house)} & --- & 0.6632 & --- & --- & --- & ---\\ \\
            & \makecell[t]{Fan et al. \\(2020)\cite{fan2020mass}} & 3D Mask-RCNN & \makecell[t]{Fudan University \\Affiliated Cancer \\ Center (in-house):\\364 DBT samples \\(289 malignant,\\75 benign)} & --- & 0.934 & 0.053 & --- & --- & --- \\ \\
            & \makecell[t]{Wichakam et al. \\(2018)\cite{wichakam2018deep}} & 3D ConvNet & \makecell[t]{(in-house):\\115 DBT volumes\\(91 malignant,\\24 normal)} & 0.72 & --- & --- & --- & --- & 0.842 \\
            \midrule
            ABUS & \makecell[t]{Lei et al. \\(2021)\cite{lei2021breast}} & Mask scoring RCNN & (private) & --- & --- & --- & $0.85\pm0.104$ & --- & --- \\ \\
            & \makecell[t]{Zhou et al. \\(2021)\cite{zhou2021multi}} & $\mathrm{C_{MS}}\mathrm{VNet_{Iter}}$ & \makecell[t]{Peking University\\People’s Hospital\\(in-house): \\900 ABUS volumes} & --- & 0.787 & --- & 0.392 & $0.778\pm0.145$ & 0.811 \\
            \midrule
            MRI & \makecell[t]{Zhou et al. \\(2019)\cite{zhou2019weakly}} & 3D DenseNet & \makecell[t]{(in-house):\\720 malignant,\\353 benign} & --- & 0.859 & --- & --- & $0.501\pm0.274$ & --- \\ \\
            & \makecell[t]{Hu et al. \\(2020)\cite{hu2020deep}} & VGG19Net & \makecell[t]{(in-house): \\728 malignant,\\ 199 benign} & --- & \makecell[t]{DEC: 0.85\\ T2w: 0.78\\ ImageFusion: 0.85\\ FeatureFusion: 0.87\\ ClassifierFusion: 0.86} & --- & --- & --- & --- \\
            \bottomrule
        \end{tabular}
    }
\end{table*}

\subsection{Overview on 2.5D CNN}
The utilization of the depth dimension and collection of images within a DBT volume is needed to utilize the entirety of the DBT information \cite{kim2017latent, zhang20192d, xiao2021classification}.
Furthermore, 2D CNN cannot preserve the between-slice information in DBT volumes \cite{liu2019anisotropic, tran2018closer, garcia2018survey, liu20173d}.
Moreover, the high complexity, potential overfitting, and small DBT dataset relative to ImageNet \cite{kim2017latent, zhang20192d, carreira2017quo} can make training a 3D CNN rather infeasible, which makes approaches by \cite{kim2017latent, liu20173d, liang2019joint} more favourable.
In the following section, alternative methods to 2D and 3D CNN approaches for utilizing the entirety of information within DBT images will be discussed.

\subsection{Current Approaches using 2.5D CNN Architecture}
Kim et al. \cite{kim2017latent} proposed a CNN for spatial feature representation and depth directional long-term recurrent learning for depth feature representation. 
A VGG16 network was used as the CNN, while LSTMs are used for depth directional long-term recurrent learning.
The model achieved an AUROC of 91.9\%.
However, a LSTM network can be difficult \cite{liu20173d} and expensive \cite{carreira2017quo} to train.

Hence, a different approach for learning 3D DBT volumes was proposed by Liu et al. \cite{liu20173d}. They proposed the 3D Anisotropic Hybrid Network (3D AH-Net).
The 3D AH-Net achieved a global dice score of 83.4\%.
As mentioned by Liu et al., challenges with directly training a 3D CNN with DBT or CT scans include (1) anisotropic voxels, (2) the higher number of features needed compared to 2D CNN, and (3) the lack of pre-trained 3D CNN models and limited training data.
Anisotropic voxels have uneven distribution of data that hinder the training of 3D CNNs, such as CT and DBT volumes having within-slice resolution greater than between-slice resolution.
This challenge of anisotropic voxels in DBT images was treated using anisotropic convolutions.
The 3D AH-Net has a feature encoder and decoder. 
The encoder extracts deep representations from the 2D image slices. 
On the other hand, the decoder, a densely connected network of anisotropic convolutions, utilizes the 3D context, while keeping the between-slices consistency.
The AH-ResNet, a 2D ResNet50 \cite{he2016deep} converted into a 3D ResNet50, was used as the backbone and encoder of the 3D AH-Net. 
A 2D Multi-Channel Global Convolutional Network (MC-GCN) was used to train the encoder parameters used in the 3D AH-NET. The parameters trained on the MC-GCN were extracted and transferred into the AH-Resnet.
The 3D AH-Net is structured in the order of AH-ResNet, decoder, then pyramid volumetric pooling.

A key challenge mentioned by Liang et al. \cite{zhang20192d} is the effective utilization of DBT data, considering as DBT data are high in resolution and vary in depth.
The training of a 3D CNN model with DBT data is computationally costly and memory intensive.
As a result, Liang et al. proposed a network containing two types of CNNs, a CNN feature extractor and CNN classifier.
The model achieved an AUROC of 97\%.
The network sequence starts with a 2D CNN feature extractor transitioning into an ensemble of three CNN classifiers. 
A classifier for DM, DBT, and DM-DBT feature map classification, where the DM-DBT feature map is a concatenation of DM and DBT feature maps.

\subsection{Overview on 3D CNN}
A 2D CNN cannot preserve the between-slice information in DBT volumes as opposed to a 3D CNN \cite{liu2019anisotropic, tran2018closer, garcia2018survey, liu20173d}.
A 3D CNN can learn the spatial information within a 2D image along with between-slice information of multiple slices.
A 3D convolutional kernel enables this characteristic, but also generates a higher number of training parameters compared to a 2D convolutional kernel.
As a result, the complexity of features that each kernels can extract increases.
A challenge arises when isotropic 3D convolutional kernels are used to learn anisotropic DBT volumes, due to variation of resolution within each anisotropic voxel along each plane \cite{liu20173d, liu2019anisotropic}.
Moreover, the quality of spatial resolution of DBT images can impact the training of 2D and 3D CNNs \cite{kopans2014digital, liu2019anisotropic, liu20173d}.
Another challenge with training 3D CNN is the limited well-curated and publicly available DBT dataset. 
Buda et al. \cite{buda2020detection} addressed this issue by curating and releasing a publicly available dataset with 22,032 reconstructed DBT volumes from 5,060 patients.

\subsection{Current Approach using 3D CNN Architecture}
Fan et al. \cite{fan2020mass} showcases the superiority of 3D over 2D deep learning methods, such as Yousefi et al. \cite{yousefi2018mass}, in learning to classify, detect, and segment tumours in DBT image slices.
Fan et al. \cite{fan2020mass} proposed a 3D-Mask-RCNN with a ResNet50 backbone.
The proposed 3D-Mask-RCNN achieved a sensitivity of 90\% at 0.83 FPs/breast for breast-based mass detection, and an AP of 93.4\% and FNR of 5.3\% for lesion segmentation.
The study concluded the proposed model, 3D-Mask-RCNN, outperforms the 2D counterparts, Mask-RCNN and Faster-RCNN.

\section{Challenges}
This section looks at the challenges faced in deep learning for breast tumour diagnosis.

\subsection{Small Dataset}
Small datasets pose a challenge to the training of deep learning models \cite{buda2020detection, gardezi2019breast, dembrower2019multi}, due to the need to familiarize the model with all possible cases to minimize classification errors.
In addition, the lack of standardized datasets makes comparing and reproducing studies difficult \cite{lee2017curated}.
A training dataset for deep learning models should provide normal mammograms, mammograms with a variety of BI-RADS 1-6, mass types, calcification, asymmetries, architectural distortion cases, and several cases in one mammogram \cite{moreira2012inbreast}.
Although techniques such as data augmentation, batch normalization, and transfer learning have been used to situate limited dataset sizes, large and well-curated datasets are still a high necessity for a well-trained model \cite{suzuki2017overview, abdelhafiz2019deep, yu2018artificial}.
Yousefi et al. \cite{yousefi2018mass} used data augmentation techniques to increase the sample size from 5,040 to 40,320 2D slices.

\subsection{Class Imbalance}
Class imbalance is another challenge for deep learning, and occurs when classes have different ratios of training data \cite{buda2018systematic, johnson2019survey, ng2016dual, ge2014handling, jeni2013facing}.
Class imbalance can cause biases in the classifier, resulting in predictions skewed towards the positive or negative class depending on the data size ratio between classes \cite{abdelhafiz2019deep}.
In addition, metrics used for measuring model performance, such as accuracy \cite{chicco2020advantages}, are susceptible to class imbalance and can affect the performance of the model.
However, there are techniques to deal with class imbalance on both the data level and classifier level.
Techniques for the data level are random undersampling and random oversampling \cite{abd2013review, johnson2019survey, buda2018systematic}, while the classifier level are cost-sensitive learning and thresholding \cite{johnson2019survey, buda2018systematic}.

\subsection{Computational Cost and Memory Constraint}
Memory constraint is an issue when dealing with training data with a large feature space, such as high resolution or high dimensional images \cite{abdelhafiz2019deep, qiu2017learning, tran2015learning, zhang20192d}.
For example, when training a 3D CNN from scratch with DBT data with large feature spaces, computational and memory cost dramatically increase.

\subsection{Image Quality Variability}
The image quality depends on the system settings and manufacturer specification of the medical screening device, while the performance of models depends on the image quality \cite{tran2021computational}.
Images from breast cancer screenings with poor resolution, sharpness, contrast, focus, or high noise can hinder the model during training, predictions, or localization of lesions \cite{ibrahim2020artificial}.

\section{Future Perspective}
Neural network models discussed in this paper have achieved promising results in classification, detection, and segmentation tasks for breast tumours. 
However, further exploration of different architectures should be made to expand possible solutions to common issues, such as costly computations and memory usage, redundancies in learning 3D data, and robustness. 
In addition, there is a need for architectures to evaluate DBT, MRI, and ABUS images with higher confidence, efficiency, and speed. 
Furthermore, state-of-the-art models require a high level of robustness and confidence to display the required level of integrity to act as a "second opinion" for radiologist.

\section{Conclusion}
Deep learning has shown significant growth in supervised-learning, and continues to grow towards better facilitating radiologists in workflow and decision-making.
This paper provided an overview on deep learning theory, the effectiveness of different deep learning architectures for breast cancer screening, and challenges faced by deep learning.
Moreover, this paper also aimed to establish a clear understanding of current progress in deep learning for breast tumour diagnosis, so future directions are easily discernible.

\bibliographystyle{IEEEtran}
\bibliography{reference}

\end{document}